\documentclass{article}

\PassOptionsToPackage{numbers,compress}{natbib}

 \usepackage[preprint]{neurips_2026}

\usepackage[utf8]{inputenc} 
\usepackage[T1]{fontenc}    
\usepackage{hyperref}       
\usepackage{url}            
\usepackage{booktabs}       
\usepackage{amsfonts}       
\usepackage{nicefrac}       
\usepackage{microtype}      
\usepackage{xcolor}         
\usepackage{comment}
\usepackage{graphicx} 
\usepackage{wrapfig}
\usepackage{amsmath}
\usepackage{amsthm}
\usepackage{pifont}
\newcommand{\cmark}{\ding{51}}
\newcommand{\xmark}{\ding{55}}
\hypersetup{
  colorlinks   = true, 
  urlcolor     = blue, 
  linkcolor    = red, 
  citecolor   = blue 
}
\title{TrustErase: Auditable Instant Machine Unlearning with Passport-Embedded Representations}

\author{
  Rutger Hendrix \\
  University of Catania \\
  Catania, Italy \\
  \texttt{rutger.hendrix@phd.unict.it} \\
  \And
  Leonardo G. Russo \\
  University of Catania \\
  Catania, Italy \\
  \texttt{leonardo.russo@studium.unict.it} \\
  \And
  Concetto Spampinato \\
  University of Catania \\
  Catania, Italy \\
  \texttt{concetto.spampinato@unict.it} \\
  \AND
  Matteo Pennisi \\
  University of Catania \\
  Catania, Italy \\
  \texttt{matteo.pennisi@unict.it} \\
  \And
  Giovanni Bellitto \\
  University of Catania \\
  Catania, Italy \\
  \texttt{giovanni.bellitto@unict.it} \\
}

\newcommand{\name}[1]{\textit{TrustErase}}

\begin{document}

\maketitle

\begin{abstract}
The demand for privacy-compliant AI has amplified the need for \textit{machine unlearning}; yet, existing retraining or distillation-based methods remain unverifiable and computationally costly.  We introduce \name{}, a verifiable, data-free unlearning framework leveraging  passport-embedded representations for instant, modular, and auditable forgetting. By treating passports as cryptographic keys within parameter-efficient adaptation layers, \name{} enables the removal of specific classes or datasets through simple deactivation, without retraining, fine-tuning, or access to the original data. A singular value based decomposition conceals passports within model weights, ensuring that unlearning actions remain transparent and provably compliant. Evaluations on MNIST, CIFAR10 and CIFAR100 show that \name{} matches or exceeds state-of-the-art benchmarks such as DELETE, L2UL, and Boundary Shrink, while operating in a strictly data-free regime. Ultimately, \name{} establishes a new paradigm for trustworthy, accountable, and instantly forgettable AI systems.

\end{abstract}

\section{Introduction}
\label{sec:intro}

The increasing adoption of large-scale pretrained models across domains such as vision, language, and multimodal reasoning~\cite{dosovitskiy2020image, carlini2021extracting, schuhmann2022laion} has magnified the challenge of machine unlearning~\cite {cao2015towards, bourtoule2021machine, nguyen2025survey}, which is the ability to remove specific knowledge or data influence from a trained model. This capability is now a regulatory and ethical requirement, mandated by frameworks such as the GDPR’s `right to be forgotten'~\cite{http://data.europa.eu/eli/reg/2016/679/oj, crawford2021excavating, acquisti2010economics}. Yet, achieving auditable and efficient unlearning remains difficult~\cite{graves2021amnesiac, ginart2019making, mahadevan2021certifiable}: models must erase information about targeted data without retraining or access to the original dataset, while maintaining performance and trust on retained knowledge~\cite{gandikota2023erasing, shan2023glaze, schramowski2023safe}.

In real-world AI deployments, retraining-based unlearning can be operationally costly or even infeasible, as it may require service interruption, repeated compliance checks, and access to training data that is no longer available. 
Nevertheless, existing methods largely rely on \textit{post-hoc} interventions. Retraining or fine-tuning approaches~\cite{thudi2022unrolling, wu2020deltagrad} such as SISA~\cite{bourtoule2021machine}, gradient-based deletion~\cite{neel2021descent}, or distillation-based unlearning~\cite{zhou2025decoupled, Hinton2015Distilling}, aim to counteract the contribution of forgotten data through additional optimization steps.   DELETE~\cite{zhou2025decoupled}, in particular, formulates a decoupled distillation framework that jointly optimizes forgetting and retention losses, achieving strong results but still requiring multiple training stages and computational overhead.  
Other lines of work, such as certified removal via influence functions~\cite{guo2019certified} or boundary-based regularization methods like L2UL~\cite{cha2024learning} and Boundary Shrink~\cite{chen2023boundary}, provide approximate or partial forgetting yet depend on the availability of training data and lack provable guarantees. Despite their effectiveness, these approaches fundamentally treat unlearning as a secondary optimization problem executed after training, even when scenarios may demand immediate compliance.\\
More recently, prompt-based unlearning~\cite{Hendrix2025PreForgettable} reframed forgetting as a native architectural property, enabling instant, data-free removal by deleting learned prompt tokens.  
However, prompt-based designs still rely on external management of prompts and offer no structural assurance that unlearning has been correctly applied once a model is shared. 
This is opposed to the growing regulatory pressure demanding mechanisms that can guarantee auditable compliance of deployed models~\cite{ai2023artificial}.  Parallel research on passport-based model verification~\cite{fan2019rethinking,oh2025seal} has shown that embedding constant matrices within network layers can bind a model’s functionality to hidden credentials. Although originally conceived for copyright protection~\cite{oh2025seal, fan2019rethinking}, this paradigm suggests a promising direction for unlearning: coupling a model’s behavior with auditable structures.

Building on these insights, we propose \textbf{\name{}}, a data-free unlearning framework with authority-mediated audit. TrustErase draws on three prior lines of work and extends them. From SEAL ~\cite{oh2025seal}, we adopt the passport-in-LoRA construction with SVD-based hiding and repurpose it for unlearning. From DELETE ~\cite{zhou2025decoupled}, we adapt the masked-distribution forgetting objective into single-stage multi-task training without a teacher. Inspired by Pre-Forgettable Models ~\cite{Hendrix2025PreForgettable}, we adopt the activate-at-inference paradigm, with passports as the modular unit. Our contributions are (i) joint multi-task training of shared LoRA factors against a passport library coexisting within one parameter space; (ii) a hypernetwork that composes atomic passports into credentials for unseen forget-set combinations without retraining or data access; and (iii) an authority-mediated audit binding each released adapter to its claimed unlearning configuration. The result is instant, modular, data-free forgetting with auditable evidence.
Experimental results demonstrate that \name{} matches or exceeds state-of-the-art unlearning methods on MNIST, CIFAR-10, and CIFAR-100, while operating in a strictly data-free regime at deployment. By shifting the computational burden to pre-optimization phase, \name{} establishes a \textit{train-once, forget-anytime} paradigm in which compliance is achieved through passport deactivation rather than post-hoc retraining.

\section{Related Work}
\label{sec:related}

\paragraph{Machine unlearning.} Machine unlearning aims to remove the influence of specific data or concepts from trained models, addressing privacy mandates such as the GDPR and the `right to be forgotten'~\cite{http://data.europa.eu/eli/reg/2016/679/oj}.
Early works explored this in traditional ML settings, where convex formulations allowed exact removal through analytical or incremental updates~\cite{ginart2019making, mahadevan2021certifiable, baumhauer2022machine}.  
However, such formulations do not generalize to deep networks~\cite{cao2015towards}, where nonlinearity and overparameterization prevent efficient inversion.  
Unlearning for deep neural networks (DNNs) has therefore focused on \textit{post-hoc} interventions.  
SISA~\cite{bourtoule2021machine} partitions the training data into independent shards, retraining only those affected by deletion requests.  
Subsequent methods accelerate unlearning via gradient tracking~\cite{thudi2022unrolling}, parameter perturbation~\cite{foster2024fast}, or Fisher-based importance weighting~\cite{golatkar2020forgetting}, yet they still depend on stored gradients or access to specific data sets.  
Other strategies, such as Boundary Shrink~\cite{chen2023boundary} and L2UL~\cite{cha2024learning}, suppress forgotten information by regularizing latent decision boundaries or selectively constraining parameter updates, but they remain data-dependent.  The recent DELETE framework~\cite{zhou2025decoupled} introduced a decoupled distillation process that explicitly separates forgetting and retention terms, enabling strong performance across class-centric tasks without access to remaining data.  
However, DELETE still requires multiple optimization phases and teacher–student distillation, limiting its use for instant or data-free unlearning. 
In contrast, Pre-Forgettable Models~\cite{Hendrix2025PreForgettable} reinterpret forgetting as an architectural property: learned prompt tokens corresponding to target classes can be removed to achieve immediate forgetting.  
While efficient and data-free, such prompt-based methods, as well as the other existing strategies, lack auditable guarantees: once models are deployed, there is no mechanism to ensure that forgotten knowledge cannot be reactivated.
\begin{table*}[t!]
\centering
\caption{Comparison of core capabilities across representative unlearning approaches. A checkmark (\cmark) indicates native support for the corresponding feature. 
}
\resizebox{\linewidth}{!}{
\begin{tabular}{lcccccc}
\hline
\textbf{Capability} & 
\textbf{L2UL}~\cite{cha2024learning} & 
\textbf{Boundary Shrink}~\cite{chen2023boundary} & 
\textbf{DELETE}~\cite{zhou2025decoupled} & 
\textbf{Pre-Forgettable}~\cite{Hendrix2025PreForgettable} & 
\textbf{\name{} (Ours)} \\
\hline
Instant unlearning            & \xmark     & \xmark     & \xmark     & \cmark & \cmark \\
Forget-data free             & \xmark     & \xmark     & \cmark & \cmark & \cmark \\
Retain-data free              & \xmark     & \xmark     & \xmark     & \cmark & \cmark \\
Multi-setting adaptability    & \cmark & \cmark & \cmark & \cmark & \cmark \\
Authority-mediated audit      & \xmark     & \xmark     & \xmark     & \xmark     & \cmark \\

\hline
\end{tabular}
}
\label{tab:framework_capabilities}
\end{table*}

\paragraph{Parameter merging.}  Task arithmetic~\cite{ilharco2022editing} 
showed that task-specific weight differences can be linearly combined in parameter space 
to compose or remove model behaviors. The broader model merging 
literature~\cite{yadav2023ties} extends this to handling parameter interference across 
many tasks. These observations motivate our hypernetwork-based~\cite{ha2017hypernetworks,mahabadi2021parameter} 
passport composition, which exploits parameter-space additivity in the low-rank subspace 
to synthesize new unlearning configurations without retraining.\\
\paragraph{Passport-based verification.} Parallel efforts in model watermarking and ownership verification have introduced \textit{passport-based} schemes~\cite{fan2019rethinking, oh2025seal}.  
These embed non-trainable matrices (passports) between LoRA weights to entangle model functionality with hidden credentials that can later be used for verification.  
SEAL~\cite{oh2025seal}, for instance, integrates constant passports into LoRA adapters and hides them through singular-value decomposition, enabling ownership proof without accuracy degradation.\\  
However, passport methods target intellectual-property protection rather than data removal; they certify ownership but not forgetting. We unify these research threads by embedding passport matrices within LoRA adapters to realize an \textit{auditable instant unlearning} framework.  Here, passports serve dual roles: as containers of task-specific knowledge and as hidden credentials enabling authority-mediated audit of the applied forgetting.  
This results in a unique approach in the unlearning landscape, as shown in Table~\ref{tab:framework_capabilities}. \name, in particular, is the only method that combines instant unlearning (removal without retraining), independence from both forget and retain data, and a built-in audit mechanism. Moreover, it supports multi-setting adaptability, not only across different forgetting granularities, such as instance-level and class-level unlearning, but also across diverse unlearning objectives. This flexibility makes \name{} applicable in a wide range of operational and regulatory scenarios.

\section{Method}
\label{sec:method}

We introduce \name{}, a framework for auditable machine unlearning that leverages passport-modulated LoRA adapters to encode multiple unlearning behaviors within a single model. Each unlearning task is associated with a fixed passport matrix that modulates shared LoRA parameters, enabling selective forgetting without retraining.

\begin{figure*}[t!]
\centering
\includegraphics[width=1\textwidth]{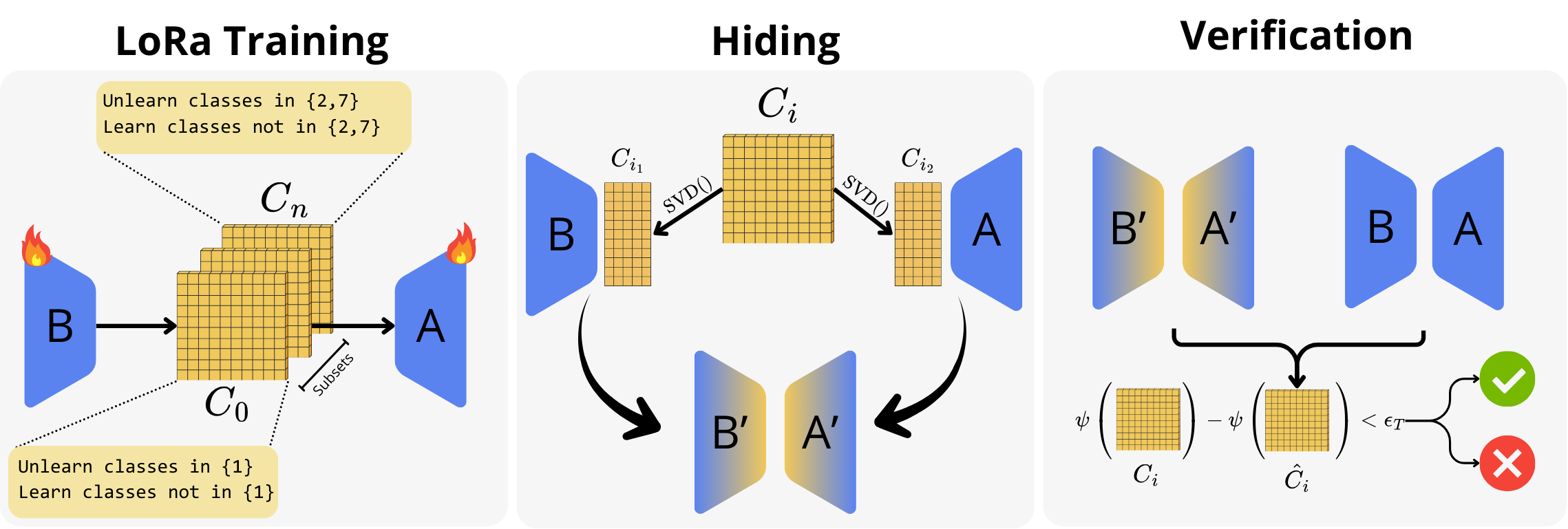}
\caption{\textbf{Overview of \name{} for Auditable Machine Unlearning.} \textit{Left:} Multi-task training with task-specific passports $\mathbf{C}_i$ inserted between shared LoRA matrices $\mathbf{A}$ and $\mathbf{B}$, enabling simultaneous optimization across multiple unlearning objectives. 
\textit{Middle:} Post-training passport hiding via SVD decomposition ($\mathbf{C}_i = \mathbf{C}_{1,i} \mathbf{C}_{2,i}$), where factors are absorbed into public weights $\mathbf{B}'_i = \mathbf{B} \mathbf{C}_{1,i}$ and $\mathbf{A}'_i = \mathbf{C}_{2,i} \mathbf{A}$. 
\textit{Right:} Audit protocol reconstructs the embedded passport $\hat{\mathbf{C}}_i = \mathbf{B}^{\dagger} \mathbf{B}'_i \mathbf{A}'_i \mathbf{A}^{\dagger}$ from public adapter and private base matrices, certifying compliance when $\|\psi(\mathbf{C}_i) - \psi(\hat{\mathbf{C}}_i)\| < \epsilon_T$.} \label{fig:method}

\end{figure*}

As illustrated in Figure~\ref{fig:method}, the framework consists of three stages: multi-task training with task-specific passports, passport hiding through matrix decomposition before model distribution, and an audit protocol that reconstructs and validates the embedded passport.

The remainder of this section details the components of the proposed framework. We first review the passport mechanism and its integration with LoRA adapters (Section~\ref{sec:background}) and then describe the unlearning formulation and training objective (Section~\ref{sec:audit_ULframework}). Next, we introduce a hypernetwork for compositional passport synthesis (Section~\ref{sec:hypernetworks}) to enable multi-class forgetting on-the-fly. Finally, we describe the audit procedure (Section~\ref{sec:verification}).

\newpage
\subsection{Background}
\label{sec:background}

\textit{Passports} were first introduced in ~\cite{fan2019rethinking} as constant matrices embedded within neural networks to enable model ownership verification.  
When integrated into training, a passport acts as a hidden credential that entangles the model parameters with a secret key: the network performs correctly only when the valid passport is present, and its accuracy degrades if the passport is removed or replaced.  
The SEAL framework~\cite{oh2025seal} extended this concept to parameter-efficient finetuning by embedding passports within the \textit{Low-Rank Adaptation} (LoRA) formulation~\cite{hu2022lora}.   In standard LoRA, the task-specific update of a pretrained weight matrix $\mathbf{W}_0 \in \mathbb{R}^{d \times k}$ is defined as
\begin{equation}
    \Delta \mathbf{W} = \mathbf{B}  \mathbf{A},
\end{equation}
with $\mathbf{B} \in \mathbb{R}^{d \times r}$, $\mathbf{A} \in \mathbb{R}^{r \times k}$, and $r \ll \min(d,k)$.
SEAL introduces a non-trainable, constant passport matrix $\mathbf{C}$ between them:
\begin{equation}\label{form:passport_seal}
    \Delta \mathbf{W} = \mathbf{B} \mathbf{C}  \mathbf{A},
\end{equation}
thereby \textit{entangling} $\mathbf{C} \in \mathbb{R}^{r \times r}$ with the trainable parameters $\mathbf{A}$ and $\mathbf{B}$.  
Before model distribution, the passport is hidden through a decomposition function $f: \mathbf{C} \mapsto (\mathbf{c}_1, \mathbf{c}_2)$ such that $\mathbf{c}_1 \times \mathbf{c}_2 = \mathbf{C}$.  
A common choice is the singular value decomposition (SVD), $\mathbf{C} = \mathbf{U} \mathbf{\Sigma} \mathbf{V}^\top$, yielding $\mathbf{c}_1 = \mathbf{U} \times \mathbf{\Sigma}^{1/2}$ and $\mathbf{c}_2 = \mathbf{\Sigma}^{1/2} \times \mathbf{V}^\top$.  
The decomposed components are then absorbed into the LoRA weights:
\begin{equation}
    \mathbf{B}' = \mathbf{B} \times \mathbf{c}_1 \quad \text{and} \quad \mathbf{A}' = \mathbf{c}_2 \times \mathbf{A},
\end{equation}
so that the distributed model $(\mathbf{B}', \mathbf{A}')$ is indistinguishable from a standard LoRA adapter, while still implicitly containing the hidden passport.

Verification relies on the fact that only the legitimate owner, who possesses the original $\mathbf{A}$ and $\mathbf{B}$, can reconstruct the passport using the Moore–Penrose pseudoinverses $\mathbf{A}^{\dagger}$ and $\mathbf{B}^{\dagger}$:
\begin{equation}
    \widehat{\mathbf{C}} = \mathbf{B}^\dagger \times \mathbf{B}' \times \mathbf{A}' \times \mathbf{A}^\dagger.
    \label{eq:reconstruction}
\end{equation}
If $\widehat{\mathbf{C}}$ statistically matches the claimed passport $\mathbf{C}$, ownership (or authorization) is verified.  
This mechanism establishes a binding between model performance and the presence of the correct passport: a property that we exploit in this work to achieve \textit{auditable instant unlearning}.

\subsection{Auditable Unlearning Framework}
\label{sec:audit_ULframework}

Building upon Section~\ref{sec:background}, \name{} extends the role of passports from ownership verification to \textit{machine unlearning}, proposing a framework where a single Vision Transformer (ViT) expresses multiple unlearning behaviors through task-specific passports modulating the LoRA updates (Eq.~\ref{form:passport_seal}). This modification is applied to the query ($Q$) and value ($V$) projection layers within each self-attention block. While $\mathbf{A}$ and $\mathbf{B}$ remain shared and trainable across all tasks,  swapping $\mathbf{C}$ allows multiple unlearning configurations to coexist within a single parameter space.

\subsubsection{Task Unlearning Definition}
\label{subsec:multitask}

Let $\mathcal{D} = \{(x_i, y_i)\}_{i=1}^{N}$ denote the complete training set. We define the space of potential forgetting configurations as the power set $\mathcal{P}(\mathcal{D})$. In practice, a finite collection of unlearning tasks is considered during training:
\begin{equation}
    \mathcal{S} = \{ S_1, S_2, \dots, S_K \} \subseteq \mathcal{P}(\mathcal{D}),
\end{equation}
where $S_k$ represents the data to forget for task $k$, and $\bar{S}_k = \mathcal{D} \setminus S_k$ the data to retain. Each task $k$ is associated with a fixed passport $\mathbf{C}_k \in \mathbb{R}^{r \times r}$, initialized randomly and kept frozen throughout training, yielding a library $\mathcal{C} = \{ \mathbf{C}_1, \dots, \mathbf{C}_K \}$ of task-specific keys.
During training, $\mathbf{A}$ and $\mathbf{B}$ are shared and updated jointly across all tasks, while each $\mathbf{C}_k$ remains fixed. The adapted forward computation for task $k$ is:
\begin{equation}
    \mathbf{h}_k = (\mathbf{W}_0 + \mathbf{B} \mathbf{C}_k \mathbf{A}) \, \mathbf{x}.
\end{equation}
By switching the active passport $\mathbf{C}_k$, the same model can forget specific classes, instances, or arbitrary data subsets without altering its base parameters or accessing the original data.

\subsubsection{Learning and Unlearning Optimization.}
\label{subsec:training}

In line with~\cite{Hendrix2025PreForgettable}, \name{} jointly optimizes $(\mathbf{A}, \mathbf{B})$ to retain general task performance while embedding the ability to selectively forget specific data subsets through the fixed passports $\{\mathbf{C}_k\}$, enabling instant unlearning without access to the original data. For each passport, the objective combines a \textit{forgetting} and a \textit{retaining} loss.

\paragraph{Forgetting Loss ($L_{\text{forget}}$).} Inspired by DELETE~\cite{zhou2025decoupled}, we adopt a masking strategy that redistributes probability mass away from forgotten classes onto retained ones. Unlike DELETE, which has post-hoc access to frozen teacher,we apply it in a single training stage using the current model's logits. Let $\mathbf{z}(x;\theta)$ be the logits from the current model and $S_k$ the indices of classes or samples to forget. We define a masking operator:
\begin{equation}
    \mathrm{Mask}_{S_k}(\mathbf{z})_i =
    \begin{cases}
        \mathbf{z}_i, & i \notin S_k,\\
        -\infty, & i \in S_k.
    \end{cases}
\end{equation}
The masked distribution is then:
\begin{equation}
    p^*(y|x,\mathbf{C}_k) =
    \mathrm{Softmax}\!\left(\mathrm{Mask}_{S_k}(\mathbf{z}(x;\theta))\right).
\end{equation}
The forgetting loss minimizes the KL divergence between this masked distribution and the current model prediction:
\begin{equation}\label{eq:forget_loss}
    L_{\text{forget}} =
    \mathrm{KL}\!\left(
        p^*(y|x,\mathbf{C}_k)
        \;\big\|\;
        q_\theta(y|x,\mathbf{C}_k)
    \right).
\end{equation}

\paragraph{Retaining Loss ($L_{\text{retain}}$).} To preserve discriminative ability on the remaining data, we apply standard cross-entropy over classes not in $S_k$:
\begin{equation}\label{eq:retain_loss}
    L_{\text{retain}} =
    - \frac{1}{|\bar{S}_k|}
    \sum_{j \notin S_k} \log p_\theta(y_j|x,\mathbf{C}_k).
\end{equation}
The overall objective for task $k$ combines forgetting and retention:
\begin{equation}
    L_k = \lambda L_{\text{forget}} + (1-\lambda)L_{\text{retain}},
\end{equation}
where $\lambda \in [0,1]$ balances the two terms. A single backward pass aggregates per-task objectives across all passports, resulting in:
\begin{equation}
    L_{\text{total}} = \frac{1}{K}\sum_{k=1}^{K} L_k.
    \label{eq:total_loss}
\end{equation}

\begin{figure}[b!]
\centering
\includegraphics[width=0.5\linewidth]{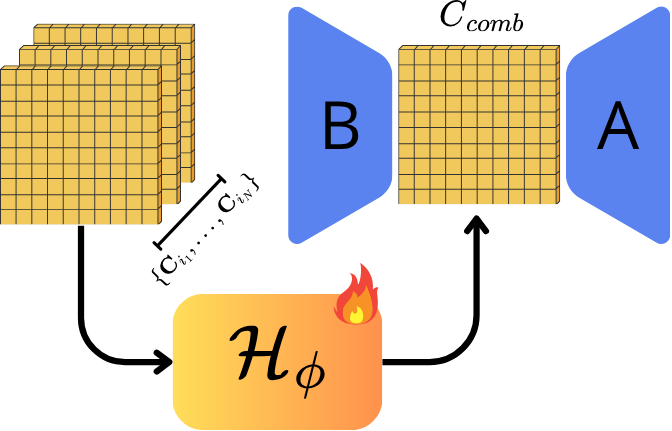}
\caption{\textbf{Hypernetwork-based Passport Composition.} Atomic passports are sampled from the power set and fed into a hypernetwork to synthesize composite passports. The hypernetwork is optimized end-to-end with frozen LoRA parameters and base passports.}
\label{fig:combiner}
\end{figure}

\subsection{Hypernetwork-based Passport Composition}
\label{sec:hypernetworks}

The framework described so far requires defining each unlearning configuration \textit{a priori}. For example, one passport may be trained to forget class 1 and another to forget class 2. If a new configuration such as forgetting both classes 1 and 2 simultaneously is needed, a dedicated passport must normally be trained for that specific combination. To avoid retraining for every possible combination, we introduce a \textit{hypernetwork-based passport synthesis} mechanism, as depicted in Fig.~\ref{fig:combiner} that learns to compose new passports from the existing atomic ones.
\paragraph{Atomic Passports.} During base training, \name{} learns a compact set of \textit{atomic passports}, each corresponding to a forgetting task  while retaining all others. For a $C$-class problem, this produces $C$ base passports $\{\mathbf{C}_1, \ldots, \mathbf{C}_C\}$, which act as compositional building blocks.

\paragraph{Hypernetwork Composition.} After training all atomic passports, a simple hypernetwork $\mathcal{H}_\phi$ is trained \textit{once} to synthesize composite passports by combining a given selection of atomic ones $\{\mathbf{C}_{i_1}, \ldots, \mathbf{C}_{i_N}\}$ associated with the classes or samples to be forgotten:
\begin{equation}
    \mathbf{C}_{\text{comb}} = \mathcal{H}_\phi(\mathbf{C}_{i_1}, \ldots, \mathbf{C}_{i_N}).
\end{equation}
During this stage, $\mathbf{A}$ and $\mathbf{B}$ remain frozen, and $\mathcal{H}_\phi$ is optimized end-to-end using the same joint objective as base training (Eq.~\ref{eq:forget_loss}, Eq.~\ref{eq:retain_loss}). The choice of a hypernetwork is motivated by the additivity of behavioral composition in parameter space, here operating in the low-rank passport space, as seen in task arithmetic and model merging literature~\cite{ilharco2022editing, yadav2023ties}. To handle variable-sized subsets, $\mathcal{H}_\phi$ leverages a Transformer encoder with padding masks; architectural details are provided in the Appendix \ref{app:supp_hypernetwork_arch}.

\subsection{Passport Verification and \texorpdfstring{$\epsilon$}{epsilon}-Tolerance}
\label{sec:verification}

The passports $\{\mathbf{C}_k\}$ encode sensitive information, as each acts as a key controlling which unlearning configuration the model performs. Distributing these matrices directly would compromise both the integrity of the unlearning mechanism and the privacy of the authorization process. Instead, we apply the SVD decomposition introduced in Section~\ref{sec:background} as a privacy mechanism, absorbing the passport factors into $(\mathbf{A}'_k, \mathbf{B}'_k)$ such that the distributed model is indistinguishable from a standard LoRA adapter. This ensures that the true passport cannot be inferred from the distributed weights alone, yet controlled audit by trusted entities remains possible through the original private matrices $(\mathbf{A}, \mathbf{B})$. Given access to the original private matrices, an estimated passport $\widehat{\mathbf{C}}_k$ can be reconstructed following equation \ref{eq:reconstruction}.

Verification then consists of checking whether $\widehat{\mathbf{C}}_k$ faithfully approximates the original passport $\mathbf{C}_k$ under a suitable distance metric $d(\cdot)$:
\begin{equation}
    \| \psi(\widehat{\mathbf{C}}_k) - \psi(\mathbf{C}_k) \| < \epsilon_T,
    \label{eq:verification}
\end{equation}
where $\psi(\cdot)$ denotes an arbitrary feature mapping and $\epsilon_T$ is a predefined tolerance accounting for numerical imprecision. If this condition holds, the adapter $(\mathbf{A}'_k, \mathbf{B}'_k)$ is certified as correctly implementing the unlearning configuration associated with $\mathbf{C}_k$, providing a lightweight and privacy-preserving form of \textit{structural verification}. To strengthen assurance, the procedure can be complemented by a \textit{functional verification} step. Here, one compares the behavior of the reconstructed model $(\mathbf{W}_0 + \mathbf{B}'_k \mathbf{A}'_k)$ with that of an ideal reference $(\mathbf{W}_0 + \mathbf{B} \mathbf{C}_k \mathbf{A})$ on a held-out set, assessing their agreement in terms of logits, intermediate activations, or task-specific metrics. When both structural and functional checks fall within the $\epsilon$-tolerance, the model is deemed to have the valid unlearning configuration.

\section{Experiments}
\label{sec:experiments}
\subsection{Experimental Setup}
\label{sec:experimental_setup}

We evaluate \name{} along four complementary dimensions: \emph{single-class unlearning}, assessing erasure of one semantic concept while preserving the others; \emph{multi-class unlearning}, testing scalability as the number of forgotten classes increases (forget-set sizes 2, 5, 10, and 20 on CIFAR-100); \emph{compositional generalization}, analyzing the hypernetwork's ability to synthesize passports for unseen class combinations; and \emph{audit robustness}, evaluating tolerance to passport perturbations. Experiments use MNIST, CIFAR-10, and CIFAR-100 on Vision Transformer backbones (ViT-Tiny, ViT-Small, ViT-Base); the framework is architecture-agnostic, and VGG-16 results are reported in Appendix~\ref{app:class_unlearning}. We compare against unlearning methods compatible with transformer architectures: \textit{Gradient Ascent}, \textit{Fisher Forgetting}~\cite{golatkar2020forgetting}, \textit{Boundary Shrink/Expand}~\cite{chen2023boundary}, \textit{Learn-to-Unlearn (L2UL)}~\cite{cha2024learning}, \textit{Saliency Unlearn (SalUn)}~\cite{fan2023salun}, \textit{Bad Teacher}~\cite{chundawat2023can}, \textit{Amnesiac}~\cite{graves2021amnesiac}, \textit{UNSIR}~\cite{tarun2023fast}, and \textit{DELETE}~\cite{zhou2025decoupled}, with \textit{Finetune}, \textit{Random Label}, \textit{Retrain} (upper bound), and \textit{Original} (lower bound) as reference points; competitor numbers are taken from their original papers. \name{} and the hypernetwork are trained with AdamW and cosine annealing for 10 epochs each, with LoRA modules applied to the query and value projections. Full training details, hyperparameter search, reconstruction attacks, instance-level forgetting, extended quantitative tables across methods, objectives, architectures, time-performance evaluation, and audit protocol use cases are detailed in the Appendix.
\paragraph{Metrics.} Following prior work~\cite{zhou2025decoupled, chen2023boundary}, we report accuracy on the forgotten test set ($Acc_{ft}$), accuracy on the retained test set ($Acc_{rt}$), and \textit{Membership Inference Attack} (MIA) success rate~\cite{shokri2017membership}, measuring residual memorization; for MIA we use a logistic regressor implementation~\cite{foster2024fast}. To compare against \textit{Pre-Forgettable Models}~\cite{Hendrix2025PreForgettable} we additionally report the average forgetting gap ($Gap_f$), defined as the distance to the target $Acc$, and its maximum deviation ($Gap_f^{\max}$). Audit robustness is quantified through \textit{Passport Verification}, which measures whether perturbed or invalid passports are correctly rejected within an $\epsilon_T$-tolerant bound.

\begin{table}[t]
\centering
\scriptsize
\setlength{\tabcolsep}{2pt}
\caption{\textbf{Single-class unlearning} performance on CIFAR-10 and CIFAR-100 for ViT-Small rank 32.}
\begin{tabular}{l|ccc|ccc}
\toprule
\textbf{Method} &
\multicolumn{3}{c|}{\textbf{CIFAR10}} &
\multicolumn{3}{c}{\textbf{CIFAR100}} \\
\midrule
 & $\text{Acc}_{ft}\downarrow$
 & $\text{Acc}_{rt}\uparrow$
 & $\text{MIA}\downarrow$
 & $\text{Acc}_{ft}\downarrow$
 & $\text{Acc}_{rt}\uparrow$
 & $\text{MIA}\downarrow$ \\
\midrule
\textit{Original Model} & 98.63 & 97.11 & 0.963  
                        & 87.59 & 90.44 & 0.816 \\
\textit{Retrain}        & 0.00 & 96.50 & 0.140  
                        & 0.00 & 85.86 & 0.308 \\
\midrule
Finetune        & 0.30 & 97.13 & 0.252
                & 41.00 & 86.10 & 0.728 \\
Random Labels   & 5.00 & 96.66 & 0.000 
                & 0.00 & 79.95 & 0.000 \\
Gradient Ascent & 0.00 & 21.42 & 0.000
                & 0.00 & 53.95 & 0.000 \\
Delete          & 0.00 & 98.52 & 0.000
                & 0.00 & 82.88 & 0.006 \\
Bad Teacher     & 32.40 & 98.24 & 0.000
                & 0.00 & 82.99 & 0.002 \\
Boundary Expand & 0.40 & 97.78 & 0.000
                & 0.00 & 82.42 & 0.994 \\
Boundary Shrink & 41.80 & 95.70 & 0.048
                & 0.00 & 82.43 & 0.038 \\
SalUn           & 8.20 & 98.19 & 0.000
                & 0.00 & 83.01 & 0.010 \\
L2UL            & 0.00 & 68.33 & 0.000 
                & 0.00 & 45.37 & 0.000 \\
UNSIR           & 0.00 & 22.76 & 0.393  
                & 0.00 & 64.77 & 0.266 \\
Amnesiac        & 0.00 & 96.51 & 0.009  
                & 0.00 & 88.07 & 0.010 \\
\midrule
\textbf{Ours}   & 0.00 & 96.82 & 0.220
                & 0.00 & 88.35 & 0.260 \\
\bottomrule
\end{tabular}
\label{tab:single_class_unlearning}
\end{table}

\begin{table*}[ht!]
\centering
\small
\caption{\textbf{Multi-class unlearning} performance on CIFAR-100 for varying numbers of forgotten classes using ViT-Small rank 32. Our method trains once to handle all subset combinations simultaneously.}
\resizebox{\textwidth}{!}{%
\begin{tabular}{l|ccc|ccc|ccc|ccc}
\toprule
\textbf{Method} &
\multicolumn{3}{c|}{\textbf{2 Classes}} &
\multicolumn{3}{c|}{\textbf{5 Classes}} &
\multicolumn{3}{c|}{\textbf{10 Classes}} &
\multicolumn{3}{c}{\textbf{20 Classes}} \\
\midrule
 & Acc$_{ft}\downarrow$ & Acc$_{rt}\uparrow$ & MIA$\downarrow$
 & Acc$_{ft}\downarrow$ & Acc$_{rt}\uparrow$ & MIA$\downarrow$
 & Acc$_{ft}\downarrow$ & Acc$_{rt}\uparrow$ & MIA$\downarrow$
 & Acc$_{ft}\downarrow$ & Acc$_{rt}\uparrow$ & MIA$\downarrow$ \\
\midrule

Finetune &
59.54 & 86.21 & 0.780 &
52.82 & 86.52 & 0.724 &
60.11 & 87.28 & 0.790 &
56.49 & 88.50 & 0.739 \\

Random Label &
4.58 & 82.73 & 0.089 &
7.67 & 75.88 & 0.061 &
4.79 & 42.57 & 0.971 &
12.55 & 29.25 & 0.941 \\

Gradient Ascent &
86.55 & 83.00 & 0.981 &
77.01 & 83.31 & 0.947 &
81.70 & 83.00 & 0.964 &
79.93 & 82.79 & 0.950 \\

Delete &
1.85 & 97.26 & 0.120 &
2.68 & 97.10 & 0.094 &
5.76 & 96.90 & 0.038 &
13.51 & 96.92 & 0.036 \\

Bad Teacher &
4.82 & 96.62 & 0.000 &
2.24 & 79.01 & 1.000 &
0.86 & 12.44 & 0.988 &
3.23 & 4.44 & 0.830 \\

Boundary Shrink &
46.51 & 83.02 & 0.702 &
46.81 & 82.44 & 0.687 &
21.51 & 54.18 & 0.841 &
17.73 & 32.35 & 0.906 \\

Boundary Expand &
19.69 & 96.74 & 0.996 &
51.47 & 94.27 & 0.012 &
58.78 & 88.64 & 0.012 &
73.48 & 87.67 & 0.012 \\
SalUn &
75.84 & 97.44 & 0.180 &
71.84 & 97.17 & 0.124 &
61.74 & 92.14 & 0.116 &
71.84 & 97.17 & 0.124 \\

L2UL &
85.55 & 83.03 & 0.977 &
75.86 & 83.26 & 0.940 &
80.91 & 83.07 & 0.964 &
79.10 & 83.11 & 0.945 \\

\midrule

Ours &
0.00 & 88.50 & 0.390 &
0.00 & 88.70 & 0.000 &
0.00 & 89.03 & 0.380 &
0.00 & 90.59 & 0.320 \\

\bottomrule
\end{tabular}}
\label{tab:multi_class_unlearning}
\end{table*}

\subsection{Unlearning performance}
\label{subsec:singleclass}
We evaluate the ability of \name{} to forget one or multiple semantic classes while preserving accuracy on the remaining ones. A single passport based LoRA model is trained once to encode all unlearning configurations jointly. Each class or subset is associated with a specific passport, and forgetting is achieved by deactivating the corresponding passports, without any retraining or access to data.
\paragraph{Single Class Unlearning.} This experiment measures how effectively the model can erase one class while retaining knowledge of the others. Table~\ref{tab:single_class_unlearning} reports results on CIFAR-10 and CIFAR-100. On CIFAR-10, \name{} reaches complete forgetting ($\text{Acc}_{ft}=0$) and $96.82\%$ test accuracy. On CIFAR-100, forgetting accuracy also reaches zero, with retained accuracy $88.35\%$ on the test set. These results are comparable to retraining based and recent unlearning methods such as DELETE and Boundary strategies, but obtained without any fine tuning or data reuse, with minimal detectable trace as illustrated by MIA scores.
\paragraph{Multi Class Unlearning.} We then extend the analysis to forgetting several classes simultaneously. Forgetting is performed by deactivating the passports of the target classes, again with no additional training. As shown in Table~\ref{tab:multi_class_unlearning}, \name{} achieves complete forgetting ($\text{Acc}_{ft}=0$) across all configurations while maintaining stable retained accuracy.  
Even when removing 20 classes on CIFAR-100, the model preserves $90.59\%$ test accuracy. Compared with Finetune, L2UL, and Random Label methods that require multiple optimization steps, \name{} attains similar forgetting quality through a single constant time operation. These results demonstrate that \name{} enables instant, auditable unlearning in both single and multi-class settings, achieving comparable (often better) performance to existing approaches while being fully data independent.

We extend our evaluation by comparing \name{} with Pre-Forgettable Models~\cite{Hendrix2025PreForgettable}, which perform modular unlearning by removing class-specific prompts. While our main experiments adopt the masked loss of equation~\ref{eq:forget_loss} to suppress forgotten logits, Pre-Forgettable Models instead enforce a uniform distribution over them. To ensure a fair comparison, we retrain \name{} using the same uniform objective, replacing the masked target distribution in equation~\ref{eq:forget_loss} with a uniform one across all classes. Table \ref{tab:uniform_objective_unlearning} summarizes single-class unlearning results on MNIST and CIFAR-10 using ViT-Tiny and ViT-Small backbones with two LoRA ranks (R8 and R32), respectively. Unlike Tables~\ref{tab:single_class_unlearning} and \ref{tab:multi_class_unlearning}, here we report the mean and standard deviation of all single-class unlearning tasks following the evaluation protocol of Pre-Forgettable Models. Across datasets and architectures, \name{} achieves forgetting gaps comparable to or smaller than those of Pre-Forgettable Models while maintaining high retained accuracy, confirming that its passport mechanism is compatible with uniform-distribution unlearning. Unlike Pre-Forgettable Models, \name{} additionally supports authority-mediated audit of the applied unlearning configuration (Sect.~\ref{sec:verification}).

\newcommand{\result}[2]{#1{\(\pm\,#2\)}}

\begin{table*}[ht]
\centering
\setlength{\tabcolsep}{3pt}
\caption{\textbf{Comparison with Pre-Forgettable Models} on MNIST and CIFAR-10 under the uniform-distribution forgetting objective. Both approaches employ LoRA adapters; results are shown for ViT-Tiny and ViT-Small backbones with two LoRA ranks (R8, R32). Reported values are mean and standard deviation averaged over all single-class unlearning tasks.}
\footnotesize{
\resizebox{\textwidth}{!}{
\begin{tabular}{l|cccc|cccc}
\toprule
\textbf{Method} &
\multicolumn{4}{c|}{\textbf{MNIST}} &
\multicolumn{4}{c}{\textbf{CIFAR10}} \\
\hline
 & $\text{Acc}_{rt}$ $\uparrow$ & $\text{Acc}_{ft}$ $\downarrow$ & $\text{Gap}_{ft}$ $\downarrow$& $\text{Gap}_f^{\max}$ $\downarrow$ &
 $\text{Acc}_{rt}$ $\uparrow$ & $\text{Acc}_{ft}$ $\downarrow$ & $\text{Gap}_{ft}$ $\downarrow$& $\text{Gap}_f^{\max}$ $\downarrow$ 
\\
\midrule
\multicolumn{9}{l}{\textbf{ViT-Tiny}} \\
\midrule
Pre-Forgettable-R8 &
\result{99.54}{0.04} & \result{11.95}{7.34} & 1.95 & 18.50 &
\result{96.17}{0.46} & \result{05.27}{6.24} & 4.73 & 12.10 \\
Pre-Forgettable-R32 &
\result{99.50}{0.04} & \result{07.46}{6.98} & 2.54 & 15.58 &
\result{96.66}{0.41} & \result{04.95}{2.62} & 5.05 & 0.80 \\

Ours-R8 &
\result{98.68}{0.16} & \result{09.45}{2.26} & 0.55 & 4.36 &
\result{88.54}{2.89} & \result{05.88}{1.60} & 4.12 & 2.00 \\
Ours-R32 &
\result{99.28}{0.09} & \result{09.47}{2.71} & 0.53 & 3.66 &
\result{93.32}{0.52} & \result{09.04}{2.10} & 0.96 & 2.20 \\
\midrule
\multicolumn{9}{l}{\textbf{ViT-Small}} \\
\midrule
Pre-Forgettable-R8 &
\result{99.63}{0.03} & \result{11.02}{8.68} & 1.02 & 14.42 &
\result{98.04}{0.23} & \result{03.46}{2.13} & 6.54 & 7.92\\
Pre-Forgettable-R32 &
\result{99.64}{0.05} & \result{07.51}{5.30} & 2.49 & 7.92 &
\result{98.32}{0.20} & \result{05.40}{4.97} & 4.60 & 8.60 \\

Ours-R8 &
\result{99.14}{0.11} & \result{10.41}{3.27} & 0.41 & 6.55 &
\result{94.94}{0.90} & \result{10.93}{4.22} & 0.93 & 10.90 \\
Ours-R32 &
\result{99.44}{0.09} & \result{10.61}{3.11} & 0.61 & 8.02 &
\result{96.32}{0.42} & \result{11.40}{2.24} & 1.40 & 5.40 \\
\bottomrule
\end{tabular}}
}
\label{tab:uniform_objective_unlearning}
\end{table*}

\begin{table}[ht]
\centering
\setlength{\tabcolsep}{4pt}
\caption{\textbf{Hypernetwork-based compositional unlearning on CIFAR-10.} Scalability across 100 seen and 100 unseen passport combinations, each composed of up to 9 atomic passports. Unseen combinations are excluded during training to evaluate generalization to novel unlearning configurations.}
\begin{tabular}{lccc}
\toprule
\textbf{Setting} & $\text{Acc}_{rt}\,\uparrow$ & $\text{Acc}_{ft}\,\downarrow$ & $\text{MIA}\,\downarrow$ \\
\midrule
Seen    & $98.92\pm 0.64$ & $0.00\pm 0.00$ & $0.250\pm 0.176$ \\
Unseen  & $96.07\pm 14.85$ & $2.22\pm 11.11$ & $0.240\pm 0.180$ \\
\bottomrule
\end{tabular}
\label{tab:hypernetwork_results}
\end{table}

We evaluate the hypernetwork described in Section~\ref{sec:hypernetworks}, which learns to synthesize composite passports by combining atomic single-class ones, enabling flexible unlearning without retraining or data access afterwards. For compositions of up to 9 unique atomic passports on CIFAR-10, the combination space grows combinatorially to over 600 possible combinations, making exhaustive evaluation infeasible. We therefore sample 100 seen and 100 unseen combinations (Table~\ref{tab:hypernetwork_results}), where the unseen combinations are deliberately excluded during training to assess generalization. For seen configurations, all 100 achieve perfect forgetting with retained accuracy above $98\%$. For unseen configurations, 96 out of 100 reach perfect forgetting with retained accuracy above $96\%$, with only 4 cases showing minimally incomplete unlearning. In practice, all combinations would be included during training. These results demonstrate that the hypernetwork generalizes well beyond its training combinations, confirming that \name{} extends its instant, data-free, and auditable unlearning to unseen tasks without additional training.

\label{sec:res_verification}

As introduced in Sect.~\ref{sec:verification}, the passport decomposition mechanism enables authority-mediated audit of unlearning configurations without exposing private credentials. Verification consists of assessing whether the reconstructed passport obtained from the released adapter remains within an $\epsilon$-tolerance of the original one, both structurally and functionally. Since random passport matrices are almost surely full-rank, honest reconstruction errors are dominated by floating-point precision, yielding relative errors of the order $10^{-6}$.
\begin{wrapfigure}{r}{0.5\linewidth}
\vspace{-5pt}
\centering
\includegraphics[width=\linewidth]{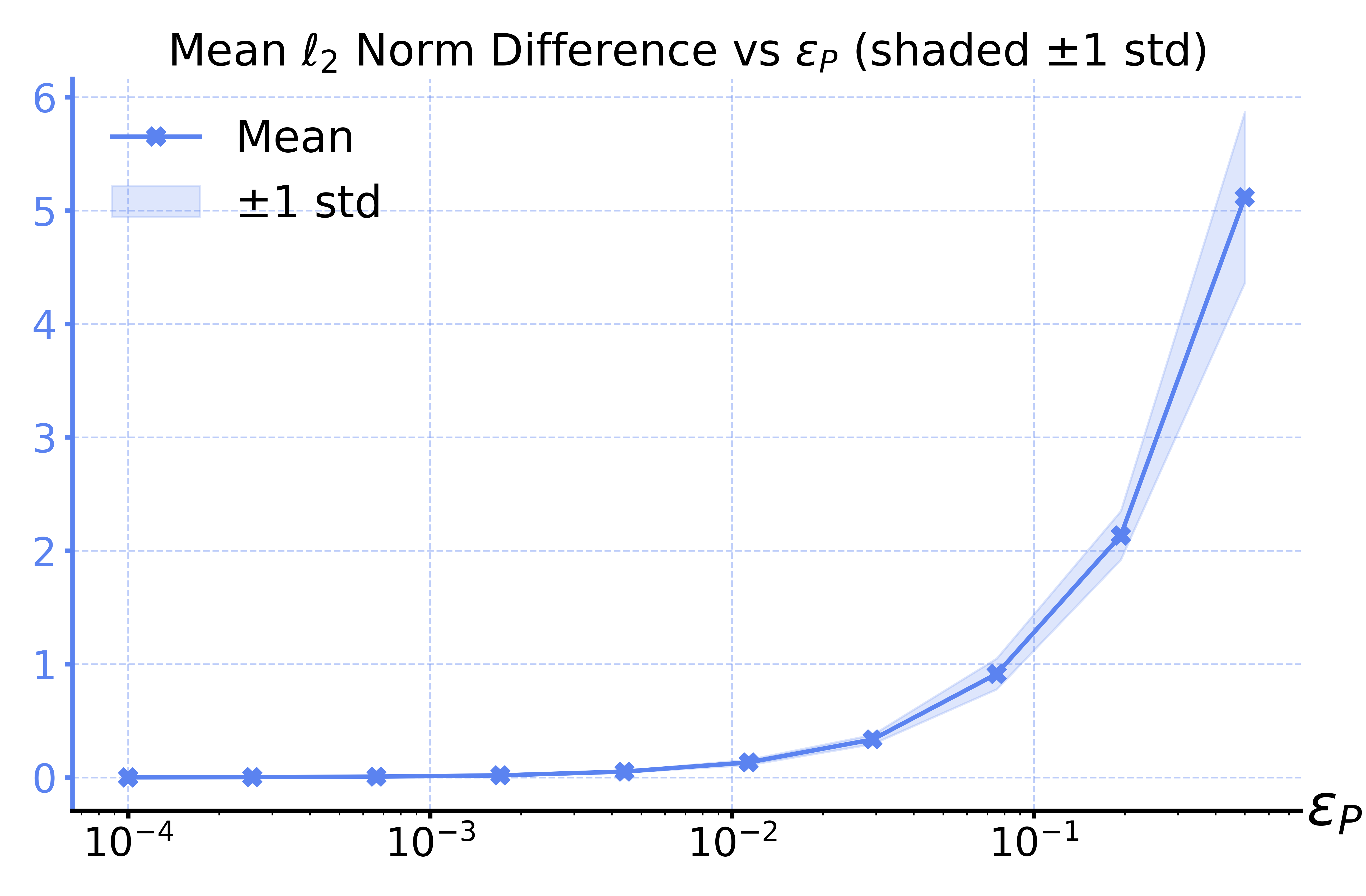}
\caption{\textbf{Sensitivity to perturbations.} 
The x-axis represents the magnitude of a random perturbation~$\epsilon_P$ added to the passport, and the y-axis shows the mean $\ell_2$ distance between the logits produced using the original and perturbed passports.}
\label{fig:epsilon_study}
\vspace{-10pt}
\end{wrapfigure}
To evaluate robustness, we perturb the passport matrix with random noise of magnitude $\epsilon_P$ and measure the resulting change in model logits (Fig.~\ref{fig:epsilon_study}). For perturbations below approximately $10^{-2}$, both the mean $\ell_{2}$ and per-logit deviations remain negligible, showing that the functional behavior of the passport-entangled LoRA update is stable within this range. Beyond this threshold, deviations grow rapidly, indicating that even small perturbations outside the tolerance window cause significant changes in model behavior. Based on this stability elbow, we set $\epsilon_T = 0.01$, which is far above numerical noise yet strictly rejects perturbations that would collapse model behavior. These results confirm that passport verification is reliable within a well-defined $\epsilon_T$ region, robust to reconstruction noise yet sensitive enough to detect incorrect or tampered passports, thereby providing a practical and auditable basis for unlearning.

\section{Discussion}
\label{sec:discussion}

\name{} reformulates machine unlearning as a pre-committed architectural property rather than a post-hoc optimization step. Authorized forgetting configurations are encoded directly into LoRA adapters during training, enabling a \textit{train-once, forget-anytime} paradigm where unlearning is performed instantly at deployment without access to user data. Passport decomposition makes unlearning configurations auditable through authority-mediated reconstruction, a property absent from existing approaches. We further demonstrate extensibility through a hypernetwork that composes atomic passports into credentials for previously unseen subset combinations, allowing the framework to scale to combinatorially large forget spaces without additional training. Experiments on MNIST, CIFAR-10, and CIFAR-100 show effective forgetting with strong retention within a single unified model, positioning \name{} as a practical mechanism for privacy-compliant AI that combines parameter efficiency, instant unlearning, and authority-mediated audit.
\paragraph{Scope and limitations.} \name{} pre-commits to atomic forget targets at training time; this is the price of the train-once, forget-anytime regime, and it rules out scenarios where the forget set cannot be defined in advance. Two extensions to already-trained models, post-hoc passport-augmented LoRA finetuning and passport distillation from existing adapters, are sketched in Appendix~\ref{app:extensions} but not empirically evaluated. Our evaluation targets the operational regime of $\leq 100$ atomic targets; extending to $\geq 1000$ targets requires addressing gradient interference between per-task losses $\{L_k\}$ in the shared low-rank subspace, for which gradient-surgery or task-grouping strategies are natural next steps (Appendix \ref{app:scaling}). Comparable modular approaches such as Pre-Forgettable Models encounter a related but distinct bottleneck: their prompt-per-class design grows the input sequence linearly with $C$, hurting inference speed and accuracy at scale. \name{} avoids this, preserving constant-time inference regardless of the number of atomic targets. At the instance level, \name{} supports bucketed forgetting via prototype-divergence partitions; per-instance passport allocation is not supported. Finally, the audit protocol is authority-mediated: it assumes an honest verification authority that holds private $(\mathbf{A}, \mathbf{B})$. Cryptographic third-party verification, for example, commitment plus zero-knowledge proof of the reconstruction identity, is left to future work.
These limitations bound the scope but not the contribution. Real-world GDPR compliance is fundamentally data-free and auditable. By the time a deletion request arrives, the data subject's records may no longer be accessible. \name{} (Table~\ref{tab:framework_capabilities}) is the only method designed for this regime, and we view it as a foundational step toward regulation-aligned machine unlearning.

\section{Acknowledgements}
Rutger Hendrix is a PhD student enrolled in the National PhD in Artificial Intelligence, XXXVIII cycle BIS, course on Health and life sciences, organized by Università Campus Bio-Medico di Roma. We acknowledge the support of the PNRR ICSC National Research Centre for High Performance Computing, Big Data and Quantum Computing (CN00000013), under the NRRP MUR program funded by the NextGenerationEU. Giovanni Bellitto, Concetto Spampinato, and Matteo Pennisi have been supported by the European Union – Next Generation EU, Mission 4 Component 2 Line 1.3, through the PNRR MUR project PE0000013 – FAIR “Future Artificial Intelligence Research” (CUP E63C22001940006).

\bibliographystyle{plainnat}
\bibliography{main}


\appendix

\clearpage
\section*{Appendix for TrustErase}
\section{Experimental Setup}
\label{app:supp_experimental_setup}

\subsection{Training details and configurations}

\textbf{Architectures.}  
Unless otherwise specified, experiments are conducted using Vision Transformers (ViT), specifically the ViT-Tiny, ViT-Small, and ViT-Base variants. All models share a depth of 12 layers and a patch size of $16 \times 16$. The architectures differ in their embedding dimensions ($d$) and number of attention heads ($h$): ViT-Tiny ($d=192, h=3$), ViT-Small ($d=384, h=6$), and ViT-Base ($d=768, h=12$). While the proposed framework is architecture-agnostic and can be applied to other backbones (e.g., CNNs), we report ViT results in the main paper for consistency with prior work; additional experiments on alternative architectures are provided in the Appendix~\ref{app:class_unlearning}.

\paragraph{Training} During training, the low-rank matrices $\mathbf{A}$ and 
$\mathbf{B}$ are shared across all tasks, while task-specific passports $\mathbf{C}_k$ 
remain fixed. A hypernetwork variant of \name{} is trained once after all atomic passports are obtained, enabling the synthesis of composite passports for arbitrary class combinations without retraining. Both stages use AdamW with weight decay $1\times10^{-4}$ and cosine annealing learning rate scheduler, with a learning rate of $1\times10^{-3}$ for base
training and $1\times10^{-4}$ for the hypernetwork, each for 10 epochs. Training uses an effective batch size of 96 via gradient accumulation. LoRA modules of rank 8, 32, and 64 are applied to the query and value projection matrices. For competing methods requiring both learning and unlearning phases, 10 unlearning epochs are applied to ensure fair comparison.

\subsection{Hyperparameter Search for Competitors}
\label{sec:supp_hyperparam_search}

To ensure a fair evaluation, all competing unlearning methods underwent a hyperparameter search. For all methods, we searched learning rates in the range $[10^{-6}, 10^{-2}]$, considering only multiples of 10, and used the configuration that achieved the best performance on CIFAR-10. For SSD and UNSIR, we searched $\alpha$ in the range $[10^{-1}, 10^{2}]$ and $\lambda$ in the range $[10^{-1}, 5]$. All competitor methods were trained using AdamW with weight decay $1\times10^{-4}$ and cosine annealing for 10 epochs, matching our setup.

\subsection{Hypernetwork Architecture}
\label{app:supp_hypernetwork_arch}

The hypernetwork $H_{\phi}$ is implemented as a task-conditioned Transformer encoder that aggregates multiple atomic passports into a single, unified representation. Given a set of $N$ atomic passports, the module synthesizes a multi-task passport $\mathcal{P}_{task}$ by integrating components selected via a task-specific mask.

Each vectorized atomic passport $P_{i} \in \mathbb{R}^{r^2}$ is projected into a shared latent space using a linear layer $W_{e} \in \mathbb{R}^{r^2 \times 768}$. A learnable $[CLS]$ token is prepended to serve as a global aggregator, and positional embeddings are added to preserve the identity of the component. The resulting input sequence $Z_{0}$ is
\[
Z_{0} = [z_{CLS}; (W_{e}P_{1} + PE_{1}); (W_{e}P_{2} + PE_{2}); \dots; (W_{e}P_{N} + PE_{N})].
\]

The fusion logic is controlled by a binary task descriptor $m \in \{0,1\}^{N}$. We utilize this descriptor to generate a padding mask $M$ applied within the Multi-Head Self-Attention (MSA) layers. For an attention head with query $Q$, key $K$, and value $V$, the attention weights are computed as
\[
\text{Attention}(Q,K,V) =
\text{softmax}\left(\frac{QK^{T}}{\sqrt{d_{k}}} + M\right)V .
\]

Here $M_{ij} = -\infty$ for all indices $j$ where $m_{j} = 0$. This ensures that the $[CLS]$ token only integrates information from the passports designated by the mask, preventing interference from inactive components.

The backbone consists of 4 encoder layers with 8 attention heads and a feed-forward expansion ratio of 4. We utilize the final hidden state of the $[CLS]$ token, $h_{CLS} \in \mathbb{R}^{768}$, as a task-conditioned bottleneck. This state is mapped back to the parameter space through a linear projection followed by a reshape operation
\[
\mathcal{P}_{task} = \text{Reshape}(W_{p} h_{CLS} + b_{p}) \in \mathbb{R}^{r \times r},
\]
recovering the passport matrix and distilling inter-task dependencies into a compact latent vector, enabling the synthesis of a unified parameterization that scales efficiently with the number of atomic components.
\label{sec:supp_extended_results}

\section{Parameter Dimensionality and Scaling Behavior}
\label{app:scaling}

\subsection{Lemma}

\paragraph{Class-independent parameter dimensionality.}
Let $\mathbf{W}_0$ be a pretrained backbone and let $\mathbf{A} \in \mathbb{R}^{r \times k}$ and $\mathbf{B} \in \mathbb{R}^{d \times r}$ denote the LoRA parameters with fixed rank $r$. Consider a classification problem with $C$ classes, where each class (or unlearning task) is associated with a fixed passport matrix $\mathbf{C}_k \in \mathbb{R}^{r \times r}$. The model prediction and training objective are
\begin{align}
f(x; \mathbf{A}, \mathbf{B}) &= (\mathbf{W}_0 + \mathbf{B} \mathbf{C}_y \mathbf{A})\, x, \\
L(\mathbf{A}, \mathbf{B}) &= \mathbb{E}_{(x,y)}\bigl[\ell(f(x; \mathbf{A}, \mathbf{B}), y)\bigr],
\end{align}
where $\ell$ is a smooth and lower-bounded loss function. Then stochastic gradient descent (SGD) converges to an $\varepsilon$-stationary point at rate $\mathcal{O}(1/\sqrt{T})$, with constants independent of the number of classes $C$.

\paragraph{Proof.}
\begin{proof}
The only trainable parameters are $(\mathbf{A}, \mathbf{B})$, whose dimension is $\mathcal{O}(r(d+k))$ and therefore does not depend on $C$. Since the passport matrices $\mathbf{C}_k$ are fixed and the loss $\ell$ is smooth, the objective $L$ has bounded-variance stochastic gradients with constants independent of $C$. Standard results  for non-convex stochastic optimization ~\cite{ghadimi2013stochastic} therefore yield
\begin{equation}
\mathbb{E}\bigl[\,\|\nabla L(\mathbf{A}_T, \mathbf{B}_T)\|^2\,\bigr] \leq \mathcal{O}(1/\sqrt{T}),
\end{equation}
which establishes convergence independently of the number of classes.
\end{proof}

\subsection{Implications and Scaling Limits}

The lemma establishes that the parameter dimensionality of the optimization problem does not grow with the number of unlearning tasks $K$. Per-step compute, however, scales linearly as $\mathcal{O}(K)$ via the per-task loss aggregation (Equation~\ref{eq:total_loss}). The lemma's smoothness assumption, combined with the gauge freedom of the passport-LoRA parameterization, means that for moderate $K$ the shared low-rank subspace has sufficient capacity to express $K$ near-orthogonal unlearning behaviors without optimization difficulty.

This regime is well-behaved within our scope. At $K \geq 1000$, the per-task gradients $\{\nabla_{(\mathbf{A}, \mathbf{B})} L_k\}$ increasingly compete for the same low-rank capacity, and the multi-task objective develops destructive interference. Concretely, when the rank-$r$ subspace cannot simultaneously accommodate $K$ task gradients, the pairwise gradients $\nabla_\mathbf{A} L_k$ and $\nabla_\mathbf{A} L_j$ for $k \neq j$ become non-orthogonal, and a fraction of pairs acquire negative cosine similarity. Under the aggregated update
\begin{equation}
\nabla_{\mathbf{A}, \mathbf{B}} L_{\text{total}} = \frac{1}{K}\sum_{k=1}^{K} \nabla_{\mathbf{A}, \mathbf{B}} L_k,
\end{equation}
the conflicting components cancel while the redundant components dilute the per-task signal, so SGD progresses suboptimally on individual $L_k$ even when $L_{\text{total}}$ continues to decrease. This is the gradient-conflict phenomenon studied in the multi-task optimization literature~\cite{sener2018mgda, yu2020pcgrad, liu2021cagrad}, and a related form of interference has been documented post-hoc when combining independently trained task-specific parameter updates~\cite{yadav2023ties}.

Two mitigation strategies are natural extensions in this regime: gradient surgery, which projects per-task gradients onto mutually orthogonal subspaces before aggregation~\cite{yu2020pcgrad}, and task grouping, which clusters passports by gradient similarity and trains each cluster against an independent $(\mathbf{A}, \mathbf{B})$ pair. Both restore the bounded-gradient-variance condition that underlies the lemma's convergence guarantee: gradient surgery by removing destructive cross-task components, task grouping by reducing the effective $K$ per parameter pair so that the rank-$r$ subspace regains sufficient capacity. We leave a full empirical study to future work.

\section{Extensions to Already-Trained Models}
\label{app:extensions}

\name{} as presented requires the passport library and the shared LoRA factors $(\mathbf{A}, \mathbf{B})$ to be jointly trained. For deployments where the target model already exists, two extension routes are possible depending on where the target knowledge resides. 

\paragraph{Backbone knowledge: passport-augmented LoRA finetuning.} If the unlearning target is encoded in the backbone itself, a passport-augmented LoRA module can be naturally attached on top of the frozen backbone and finetuned with the multi-task objective (Equation~\ref{eq:total_loss}), with the target classes included in the per-task losses $\{L_k\}$. 

\paragraph{LoRA-encoded knowledge: passport distillation.} If the target knowledge already lives in an existing LoRA adapter $(\mathbf{A}_{\text{src}}, \mathbf{B}_{\text{src}})$, that adapter can be distilled into a passport-augmented LoRA via supervised passport distillation. Concretely, a fresh passport library $\{\mathbf{C}_k\}$ and shared factors $(\mathbf{A}, \mathbf{B})$ are trained to match the source adapter's outputs on the retain set, while applying the masked forgetting loss (Equation~\ref{eq:forget_loss}) on the forget set.

Both routes inherit the pre-commitment requirement: the atomic forget targets must be specified at finetune or distillation time. Empirical evaluation of these extensions is left to future work.

\section{Class Unlearning}
\label{app:class_unlearning}

\paragraph{Cifar20 evaluation.}
Here we extend our evaluation to CIFAR-20. Under the setup of competing methods, Table~\ref{tab:cifar20_class_unlearning} demonstrates that our approach achieves perfect forgetting in single-class scenarios while preserving strong retain performance and privacy guarantees close to retraining baselines. Table~\ref{tab:preforgettable_ours_cifar20} further shows that, under a uniform-distribution forgetting objective, we achieve on-par forget accuracy compared to Pre-Forgettable methods while exhibiting substantially smaller worst-case forgetting gaps, indicating more consistent and reliable forgetting behavior.

\begin{table*}[ht]
\centering
\resizebox{\textwidth}{!}{%
\begin{tabular}{llcccccccc}
\toprule
Class & metric & Original Model & Retrain & Finetune & Bad Teacher & UNSIR & Amnesiac & SSD & Ours \\
\midrule
Veh2 & $\mathcal{D}_r$ & 95.73$\pm$0.00 & 94.85$\pm$0.13 & 87.75$\pm$1.64 & 93.59$\pm$0.3 & 93.56$\pm$0.32 & 93.88$\pm$0.15 & 93.12$\pm$0.00 & 94.99$\pm$0.08 \\
 & $\mathcal{D}_f$ & 95.22$\pm$0.00 & 0.00$\pm$0.00 & 0.04$\pm$0.12 & 4.88$\pm$4.12 & 70.31$\pm$5.03 & 0.00$\pm$0.00 & 0.00$\pm$0.00 & 0.00$\pm$0.00 \\
 & MIA & 84.04$\pm$0.00 & 22.96$\pm$0.03 & 38.15$\pm$0.08 & 0.02$\pm$0.00 & 48.98$\pm$0.07 & 1.20$\pm$0.00 & 7.04$\pm$0.00 & 23.74$\pm$11.36 \\
\midrule
veg & $\mathcal{D}_r$ & 95.59$\pm$0.00 & 94.54$\pm$0.21 & 87.09$\pm$1.24 & 92.92$\pm$0.51 & 93.25$\pm$0.35 & 93.29$\pm$0.41 & 95.71$\pm$0.00 & 94.78$\pm$0.14 \\
 & $\mathcal{D}_f$ & 97.57$\pm$0.00 & 0.00$\pm$0.00 & 0.30$\pm$0.29 & 8.28$\pm$6.79 & 89.02$\pm$2.41 & 0.02$\pm$0.07 & 0.00$\pm$0.00 & 0.00$\pm$0.00 \\
 & MIA & 91.32$\pm$0.00 & 4.41$\pm$0.01 & 14.72$\pm$0.05 & 0.02$\pm$0.00 & 58.67$\pm$0.04 & 1.02$\pm$0.00 & 1.88$\pm$0.00 & 3.08$\pm$1.37 \\
\bottomrule
\end{tabular}
}
\caption{\textbf{Class unlearning results on CIFAR-20 using ViT-Base Rank 32}, evaluating simultaneous forgetting of two classes from a single trained model. Results for baseline methods are reported from SSD~\cite{foster2024fast}.}

\label{tab:cifar20_class_unlearning}
\end{table*}

\begin{table}[ht]
\centering
\small
\setlength{\tabcolsep}{3pt}
\caption{\textbf{Comparison with Pre-Forgettable Models on CIFAR-20 under the uniform-distribution forgetting objective.} All methods employ LoRA (ranks 32/64) with ViT-Small and ViT-Base backbones. Atomic passports are trained once, with one passport per class, and the reported results are averaged over
all classes.}
\begin{tabular}{lcccc}
\hline
\textbf{Method} & $\text{Acc}_{rt}\,\uparrow$ & $\text{Acc}_{ft}\,\downarrow$ & $\text{GAP}_{ft}\,\downarrow$ & $\text{GAP}_{f}^{\max}\,\downarrow$ \\
\hline
\multicolumn{5}{l}{\textbf{ViT-Small}} \\
\hline
Pre-Fgt-R32 & \result{94.46}{0.19} & \result{6.85}{4.14} & 1.85 & 12.00 \\
Pre-Fgt-R64 & \result{94.56}{0.17} & \result{5.70}{2.74} & 0.70 & 5.00 \\
Ours-R32    & \result{75.59}{2.93} & \result{1.90}{1.57} & 3.10 & 5.00 \\
Ours-R64    & \result{83.93}{0.86} & \result{2.10}{1.37} & 2.90 & 4.60 \\
\hline
\multicolumn{5}{l}{\textbf{ViT-Base}} \\
\hline
Pre-Fgt-R32 & \result{95.41}{0.16} & \result{7.03}{4.67} & 2.03 & 14.20 \\
Pre-Fgt-R64 & \result{95.65}{0.12} & \result{5.34}{5.46} & 0.34  & 21.00 \\
Ours-R32    & \result{87.68}{1.33} & \result{5.29}{3.01} & 0.29 & 11.20 \\
Ours-R64    & \result{92.25}{0.44} & \result{6.09}{1.99} & 1.09 & 5.20 \\
\hline
\end{tabular}
\label{tab:preforgettable_ours_cifar20}
\end{table}

\paragraph{Convolutional Neural Network evaluation.}
Although our method builds on LoRA, originally introduced for Transformers, it is not inherently Transformer-specific and can be extended to CNN’s. To demonstrate backbone agnosticism, we evaluate TrustErase on a VGG16 by inserting a 128 rank LoRA into the last 2 layers before the head and train for 10 epochs. As shown in Table \ref{tab:single_class_unlearning_vgg_16} the top competitor DELETE exhibits unstable retention below 80\% and residual memorization, while TrustErase maintains high retention (88.11\%) and achieves perfect forgetting without a trace. These results confirm that the mechanism transfers effectively beyond Transformers.

\begin{table}[t]
\centering
\scriptsize
\setlength{\tabcolsep}{2pt}
\caption{\textbf{Single-class unlearning} performance on CIFAR-10 using VGG-16.}
\begin{tabular}{l|ccc}
\toprule
\textbf{Method} &
$\text{Acc}_{ft}\downarrow$ &
$\text{Acc}_{rt}\uparrow$ &
$\text{MIA}\downarrow$ \\
\midrule
\textit{Original Model} & 90.20 & 90.63 & 0.914 \\
\textit{Retrain}        & 0.00& 90.92 & 0.298\\
\midrule
Finetune        & 75.90 & 90.76 & 0.204 \\
Random Labels   & 4.50 & 60.20 & 0.082 \\
Gradient Ascent & 23.20 & 48.36 & 0.516 \\
Delete          & 0.80 & 79.96 & 0.017 \\
Bad Teacher     & 3.80 & 63.82 & 0.062 \\
Boundary Expand & 19.90 & 52.91 & 0.240 \\
Boundary Shrink & 0.00 & 42.74 & 0.928 \\
SalUn           & 26.80 & 63.62 & 0.400 \\
L2UL            & 0.00 & 11.11 & 0.000 \\
\midrule
\textbf{Ours}   & 0.00 & 88.11 & 0.000 \\
\bottomrule
\end{tabular}
\label{tab:single_class_unlearning_vgg_16}
\end{table}

\section{Instance-Level unlearning}
\label{sec:supp_instance_forgetting}

We additionally demonstrate our framework in the challenging \emph{instance-level unlearning} setting, where the goal is to selectively forget a small subset of the training data while retaining performance on the remaining samples. For this, we construct a forget set by randomly selecting 1\% of the training data from each class of CIFAR-10 and MNIST. These samples are assigned to a dedicated `forget' passport, which is trained jointly with the retain-all passport, following our main training setup.

To encourage the model to actively diverge from the forgotten instances while preserving structure in the retained feature space, we introduce a prototype-inspired divergence loss. Let $f(x) \in \mathbb{R}^D$ denote the backbone features of an input $x$, and let $\mathcal{D}_{\text{retain}}$ be the set of retained samples in a batch. We compute class prototypes
\begin{equation}
\mathbf{p}_c = \frac{1}{|\mathcal{D}_{\text{retain}}^{(c)}|} 
\sum_{x_i \in \mathcal{D}_{\text{retain}}^{(c)}} 
\frac{f(x_i)}{\|f(x_i)\|_2},
\end{equation}
where $\mathcal{D}_{\text{retain}}^{(c)}$ denotes retained samples of class $c$.  
For each feature from the forget set $f(x_j)$, we apply a prototype repulsion objective that encourages its representation to deviate from the class prototypes:
\begin{equation}
\mathcal{L}_{\text{div}} 
= \frac{1}{|\mathcal{D}_{\text{forget}}|}
\sum_{x_j \in \mathcal{D}_{\text{forget}}}
\exp\!\left(
-\frac{1}{\tau}
\min_{c} \| f(x_j) - \mathbf{p}_c \|_2^2
\right),
\end{equation}
where $\tau$ is a temperature hyperparameter.  
This loss softly repels forgotten samples from all class prototypes, thereby promoting instance-level forgetting in the feature space.

Table~\ref{tab:instance_level_results} reports the instance-level forgetting performance on CIFAR-10 and MNIST using ViT-small rank 64. All metrics indicate effective forgetting while preserving accuracy on the remaining data.

\begin{table}[ht]
\centering
\setlength{\tabcolsep}{5pt} 
\caption{\textbf{Instance-level forgetting} results on CIFAR-10 and MNIST.}
\begin{tabular}{lccc}
\toprule
\textbf{Dataset} & $\text{Acc}_{ft}\,\downarrow$ & $\text{Acc}_{rt}\,\uparrow$ & $\text{MIA}\,\downarrow$ \\
\midrule
CIFAR-10 & $0.00 \pm 0.00$ & $94.65 \pm 0.00$ & $0.233$ \\
MNIST    & $0.00 \pm 0.00$ & $94.94 \pm 0.00$ & $0.439$ \\
\bottomrule
\end{tabular}
\label{tab:instance_level_results}
\end{table}

\section{Backbone Statistics Unlearning}
Beyond masked and uniform output objectives, we also demonstrate that the proposed method can be adapted to other unlearning objectives, such as those derived from backbone statistics. Specifically, we minimize the KL divergence between the model’s outputs and those produced by the frozen backbone combined with a randomly initialized classifier, effectively approximating the pre-learning state.
Table~\ref{tab:ours_backbone_stats} summarizes the results obtained using these backbone-based statistics. Here, the unlearning gap is defined as the distance to the pre-learning classifier accuracy, rather than the chance level, as used previously for the uniform objective. 
The consistently small accuracy gaps across varying forgetting scenarios indicate that the backbone statistics objective provides a reliable and predictable unlearning target. By anchoring to the original backbone's feature space, this approach preserves useful representations from the pretrained model.  Overall, this method provides a practical and stable alternative to uniform objectives and demonstrates our framework's capability to accommodate diverse unlearning objectives. 

\begin{table*}[ht]
\centering
\setlength{\tabcolsep}{3pt}
\caption{\textbf{Single-class unlearning performance using backbone statistics} with LoRA ranks R8 and R32 on MNIST and CIFAR-10 across various ViT backbones. Atomic passports are trained once, with one passport per class, and the reported results are averaged over all classes.}
\resizebox{\textwidth}{!}{%
\begin{tabular}{l|cccc|cccc}
\toprule
\textbf{Method} &
\multicolumn{4}{c|}{\textbf{MNIST}} &
\multicolumn{4}{c}{\textbf{CIFAR10}} \\
\midrule
\midrule
& 
    $\text{Acc}_{rt}\,\uparrow$ & $\text{Acc}_{ft}\,\downarrow$ & $\text{GAP}_{ft}\,\downarrow$ & $\text{GAP}_{f}^{\max}\,\downarrow$ &
    $\text{Acc}_{rt}\,\uparrow$ & $\text{Acc}_{ft}\,\downarrow$ & $\text{GAP}_{ft}\,\downarrow$ & $\text{GAP}_{f}^{\max}\,\downarrow$ \\
\midrule
\midrule
\multicolumn{9}{l}{\textbf{ViT-Tiny}} \\
\midrule
Ours-R8  &
\result{98.52}{0.20} & \result{7.74}{12.46} & 1.46 & 5.04 &
\result{88.29}{2.02} & \result{8.36}{12.65} & 1.14 & 3.80 \\
Ours-R32 &
\result{99.24}{0.08} & \result{8.13}{13.31} & 1.08 & 3.40 &
\result{92.87}{0.34} & \result{8.63}{13.05} & 1.53 & 3.20 \\
\midrule
\multicolumn{9}{l}{\textbf{ViT-Small}} \\
\midrule
Ours-R8  &
\result{99.07}{0.13} & \result{1.08}{25.03} & 1.64 & 6.00 &
\result{94.46}{1.15} & \result{7.95}{20.95} & 1.93 & 9.90 \\
Ours-R32 &
\result{99.38}{0.09} & \result{10.57}{25.04} & 1.89 & 6.14 &
\result{95.58}{0.33} & \result{7.93}{21.33} & 2.35 & 12.10 \\
\bottomrule
\end{tabular}
}
\label{tab:ours_backbone_stats}
\end{table*}

\section{Inference and Training Time Analysis}
\label{sec:supp_runtime}

All runtime measurements were conducted on a NVIDIA H100 GPU for CIFAR-10 single-class unlearning. Across all competing unlearning methods, the average total unlearning time is 18.9 minutes, with a standard deviation of 8.7 minutes. 
The two slowest competitors are L2UL (32.1 minutes) and Finetune (26.7 minutes).

For a realistic comparison, we consider \name{} trained for 11 initial tasks (one for each class, plus an all-learning task). The total training time for our method is 62.53 minutes, which is about 32 minutes slower on average than the total training and unlearning time of competing methods. However, in return, our approach provides instant unlearning for predefined forget sets without requiring access to the training data or additional training time.
If a forget set is not included in the predefined collection, the second hypernetwork training phase, which in the worst case learns all possible atomic-passport combinations, takes at most 1 day. This phase can be carried out after initial model deployment.

\section{Reconstruction Attack Analysis}
\label{sec:supp_reconstruction_attack}

We examine whether the shared LoRA bottleneck $C$ can be reconstructed from publicly released decomposed adapters. The setting assumes a frozen ViT backbone instrumented with LoRA adapters on all Q/V projections. For each layer $\ell$, the victim applies a low-rank update
\begin{equation}
\Delta W_\ell = \frac{\alpha}{r} B_\ell C A_\ell ,
\end{equation}
where the per-layer matrices $\{A_\ell,B_\ell\}$ and the shared $C\in\mathbb{R}^{r\times r}$ are private. Before release, the victim deterministically decomposes $C=C_1C_2$ and publishes
\begin{equation}
A'_\ell = C_2 A_\ell, \qquad B'_\ell = B_\ell C_1,
\end{equation}
so that the public adapter appears as a standard LoRA update $\Delta W_\ell = \frac{\alpha}{r} B'_\ell A'_\ell$.

The attacker instantiates its own factors $\{\hat{A}_\ell,\hat{B}_\ell\}$ and a shared $\hat{C}$, inducing
\begin{equation}
\Delta \hat{W}_\ell = \frac{\alpha}{r} \hat{B}_\ell \hat{C} \hat{A}_\ell .
\end{equation}
To remain consistent with the public decomposition, $\hat{C}$ is factored as $\hat{C}=\hat{C}_1\hat{C}_2$, defining $\hat{A}'_\ell=\hat{C}_2\hat{A}_\ell$ and $\hat{B}'_\ell=\hat{B}_\ell\hat{C}_1$. The attacker optimizes a combined objective
\begin{equation}
L_{\text{tot}} = L_{\text{cls}} + \lambda L_{\text{rec}},
\end{equation}
where the supervised loss is
\begin{equation}
L_{\text{cls}} =
\frac{1}{|D|}
\sum_{(x,y)\in D}
\mathrm{CE}\big(
f(x;\,W_\ell+\Delta\hat{W}_\ell),\; y
\big),
\end{equation}
and the reconstruction loss enforces agreement with the released public factors:
\begin{equation}
L_{\text{rec}} =
\sum_{\ell}
\left(
\|\hat{A}'_\ell - A'_\ell\|_F^2
+
\|\hat{B}'_\ell - B'_\ell\|_F^2
\right).
\end{equation}

Because the LoRA parameterization admits an $r\times r$ gauge freedom (basis changes in $\hat{C}$ that can be absorbed by $\hat{A}_\ell$ and $\hat{B}_\ell$ without altering $\Delta\hat{W}_\ell$) the inverse problem is highly underdetermined. As a consequence, many attacker optima match the public factors only loosely while remaining incompatible with the victim’s private $\{A_\ell,B_\ell\}$.

Evaluation uses both parameter-level and function-level metrics. Factor reconstruction errors (FR-A, FR-B) and update-agreement error $E_{\Delta W}$ quantify alignment with public factors, while functional fidelity is measured by substituting $\hat{C}$ into the victim model and comparing predictions, accuracy, and logit similarity with the true-$C$ model.

We optimized the attack objective for 20 epochs using AdamW with a learning rate of $10^{-3}$, weight decay of $10^{-4}$, and a cosine annealing scheduler. Despite this configuration, empirically, reconstruction fails. Errors remain large (FR-A$\approx0.976$, FR-B$\approx0.952$, $E_{\Delta W}\approx0.947$), indicating poor alignment with the released factors. More critically, transplanting $\hat{C}$ into the victim collapses performance: $\text{Acc}(\hat{C})=0.0955$ versus $\text{Acc}(C)=0.8712$, with negative logit cosine similarity and $\mathrm{KL}\approx 3.70$. Although the attacker achieves low loss on its own model, optimization exploits gauge degrees of freedom rather than recovering the true bottleneck.

Under this reconstruction threat model, the shared LoRA bottleneck $C$ is not identifiable. The decomposition-based release reveals insufficient structure, and the attacker’s optimization converges to degenerate solutions that fail both in parameter space and in function space. The passport matrix $C$ therefore remains unrecoverable.

\section{Audit Protocol and Use Cases}
\label{sec:supp_verification}

\begin{wrapfigure}{r}{0.5\linewidth}
    \centering
    \includegraphics[width=\linewidth]{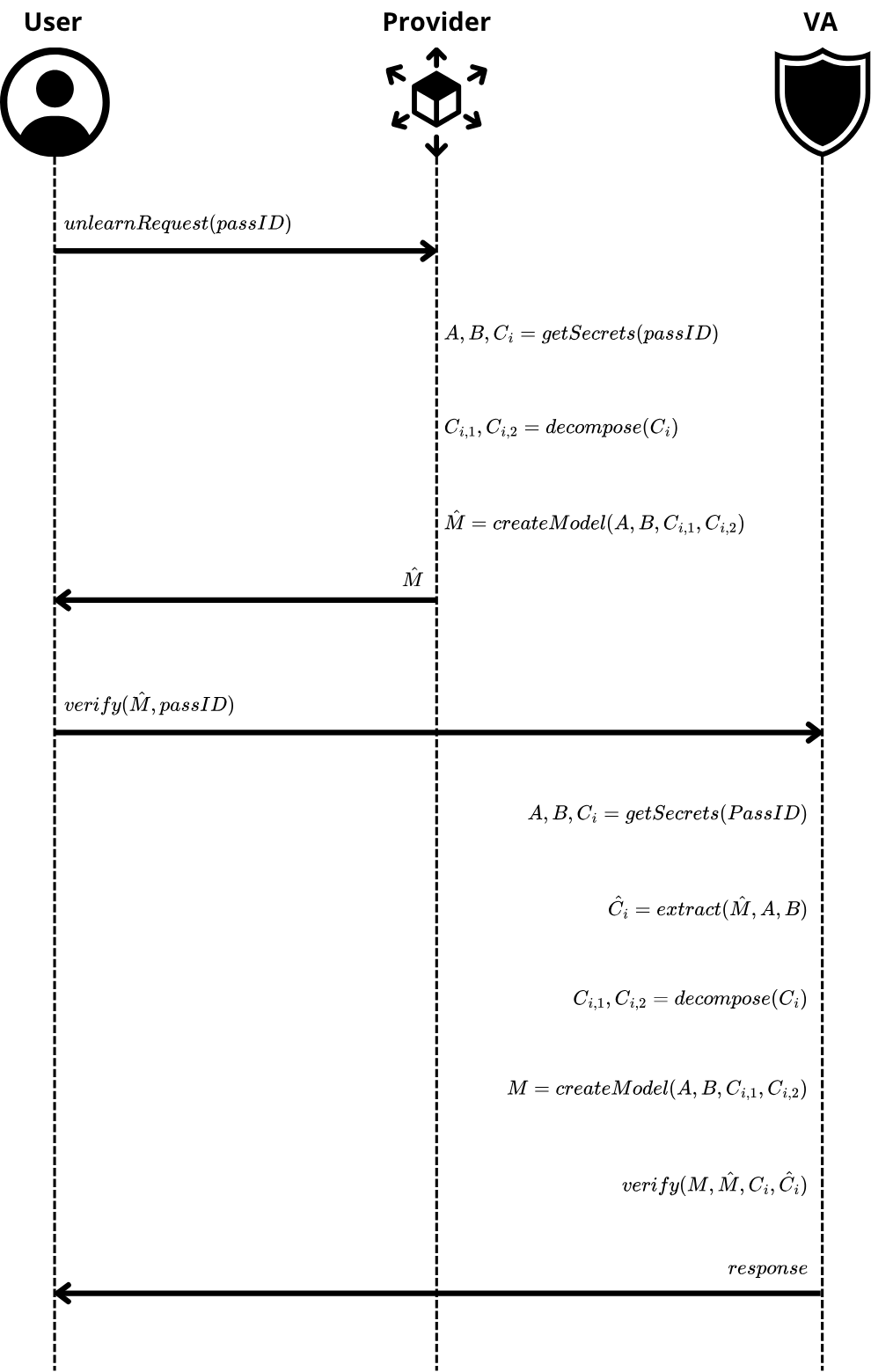}
    \caption{\textbf{Overview of the proposed verification protocol for auditable unlearning.} The protocol follows a trust-but-verify paradigm involving three entities: a Verification Authority (VA) that conducts initial training and distributes secrets, trusted providers that perform model serving using designated passports, and users who request specific unlearning tasks. The VA validates both the structural correctness of the passport used and the functional behavior of the resulting model, ensuring that the delivered model faithfully reflects the user’s intended unlearning request.}
    \label{fig:verification_protocol}
\end{wrapfigure}

To guarantee the integrity of the unlearning process and foster trust in deployed models, we design an audit protocol. The objective of this protocol is to provide users with a formal assurance that the model they receive from a service provider correctly incorporates the specific unlearning task they requested.
The protocol operates under a `trust-but-verify' paradigm, as illustrated in Figure~\ref{fig:verification_protocol}, involving three main entities:

\begin{itemize}
    \item \textbf{Verification Authority (VA):} The central entity that performs the initial training and secret distribution. It acts as the root of trust for certifying compliance.
    \item \textbf{Trusted Providers:} Entities initially entrusted with all secrets (base matrices and passports) to handle model serving. They are assumed to be honest-but-verifiable.
    \item \textbf{Users:} Clients who request specific unlearning tasks (e.g., forgetting a specific class) and verify the outcomes through the VA.
\end{itemize}

Beyond verifying the correctness of the unlearning operation itself, the protocol ensures that the resulting model exhibits functional behavior consistent with the intended task. Thus, verification encompasses both \textit{structural validation}, checking that the correct passport was used, and \textit{functional validation}, which assesses whether the produced model behaves similarly to one built with the user's original task specification.

\paragraph{Secrets Distribution.}
After central training, the VA provisions the base LoRA matrices $A,B$ together with the full set of original passports $\{C_k\}$ to a set of Trusted Providers. Each user $U_i$ is securely issued their ground-truth passport $C_i$, which serves as a non-transferable receipt for task $i$. Providers are assumed to be honest-but-verifiable: they store the secrets and serve models, but the VA remains responsible for certifying that the correct unlearning configuration has been applied.

\paragraph{Verification Workflow.}
The verification process is initiated by the User. After receiving the unlearned model for task $i$, the user contacts the VA and submits both the received model parameters (specifically the decomposed adapters $A'_i, B'_i$) and their ground-truth passport $C_i$. Upon receipt, the VA independently audits the submitted model. Leveraging the private base matrices $A$ and $B$ (which are retained securely by the VA), it extracts the effective passport $\hat{C}_i$ embedded within the model weights via pseudo-inverse reconstruction:
\begin{equation}
\hat{C}_i = B^{\dagger} B'_i A'_i A^{\dagger}.
\tag{\ref{eq:reconstruction}}
\end{equation}

Simultaneously, the VA instantiates the reference `honest' model by combining the secure base matrices with the user's verified ground-truth passport $C_i$. The VA now holds two distinct objects: the ideal passport $C_i$ and the passport $\hat{C}_i$ extracted from the deployed model. Verification consists of checking whether $\hat{C}_i$ is a faithful approximation of $C_i$ within a predefined tolerance:
\begin{equation}
    \| \psi(\widehat{\mathbf{C}}_i) - \psi(\mathbf{C}_i) \| < \epsilon_T,
    \tag{\ref{eq:verification}}
\end{equation}
where $\psi(\cdot)$ denotes an arbitrary feature mapping and $\epsilon_T$ is a predefined tolerance accounting for numerical imprecision. If the condition holds, the VA certifies that the provider has correctly applied the requested unlearning task.

As discussed in Section ~\ref{sec:supp_reconstruction_attack}, structural similarity alone may not suffice; adversarial reconstruction attacks can exploit small deviations. Therefore, the VA complements the structural check with a \emph{functional verification} step, comparing the behavior of the served model $(W_0 + B\hat{C}_iA)$ against a reference built with $C_i$. Agreement in logits or activations within $\epsilon_T$ ensures that the unlearning task has been faithfully executed.

\paragraph{Use Cases.}
This protocol supports multiple deployment scenarios: (i) regulatory audits, where third parties require verifiable evidence of compliance with `right-to-be-forgotten' requests; (ii) multi-tenant services, where different users demand distinct unlearning configurations; and (iii) adversarial settings, where providers must prove that no unauthorized passport substitution has occurred. In all cases, the combination of structural and functional verification provides lightweight yet auditable guarantees of trustworthy unlearning.
\clearpage



\end{document}